\documentclass[prd,superscriptaddress,onecolumn,showkeys]{revtex4}
\usepackage{eurosym}
\usepackage{amsfonts}
\usepackage{array}
\usepackage{amsthm}
\usepackage{bm}
\usepackage{palatino}
\usepackage{mathpazo}
\usepackage{amssymb}
\usepackage{eurosym}
\usepackage{amsmath}
\usepackage{epsfig}
\usepackage{graphics}
\usepackage{changes}
\usepackage{color}
\usepackage{graphicx}
\usepackage[colorlinks=true,
            linkcolor=blue,
           urlcolor=black,
           citecolor=black]{hyperref}

\def\be{\begin{equation}}
\def\ee{\end{equation}}
\def\beq{\begin{eqnarray}}
\def\eeq{\end{eqnarray}}

\def\bes{\begin{eqnarray}}
\def\ees{\end{eqnarray}}

\begin{document}
\title{\textbf{Tunneling Analysis of Regular Black Holes with Cosmic Strings-Like Solution in Newman-Janis Algorithm}}

\author{Riasat Ali}
\email{riasatyasin@gmail.com}
\affiliation{Department of Mathematics, Shanghai University,
Shanghai, 200444, Shanghai, People’s Republic of China}

\author{Rimsha Babar}
\email{rimsha.babar10@gmail.com}
\affiliation{Division of Science and Technology, University of Education, Township, Lahore-54590, Pakistan}

\author{Muhammad Asgher}
\email{m.asgher145@gmail.com}
\affiliation{Department of Mathematics, The Islamia
University of Bahawalpur, Bahawalpur-63100, Pakistan}

\author{Xia Tie-Cheng}
\email{xiatc@shu.edu.cn}
\affiliation{Department of Mathematics, Shanghai University,
Shanghai, 200444, Shanghai, People’s Republic of China}
\begin{abstract}
We consider the regular black holes solution with cosmic strings(RBHCS) in the rotation parameter by assuming the Newman-Janis method.
After this, we study thermodynamical property (i.e., Hawking temperature $T_H$)
for the RBHCS in the presence of spin parameter.
Moreover, we study the graphical interpretation of Hawking temperature with event
horizon to check the physical and stable form of RBHCS under the effect of Newman-Janis algorithm. We graphically show that the RBHCS in the context of Newman-Janis algorithm are colder than
the Schwarzschild black hole. Furthermore, we investigate the quantum corrected
temperature for RBHCS in Newman-Janis method by incorporating generalized uncertainty principle. We have also analyzed
the graphical interpretation of corrected temperature $T'_{H}$ versus $r_{+}$ and study the
stable condition of RBHCS in Newman-Janis method in the presence of gravity parameter effects. Finally, the corrected entropy for RBHCS with rotation parameter is analyzed. 
\end{abstract}


\date{\today}
\maketitle
\section{Introduction}John Michell initially predicted the concept of a black hole (BH) in $1783$.
Subsequently, the advanced idea of BHs was established in $1915$ with
Einstein's general theory of relativity (GR) \cite{1a}.
The mystery of the present is preserved through the BH physics, not only within
the exploration of gravitational waves \cite{1} as well as by the more basic level
of BH physics (i.e., information paradox and entropy) \cite{2}. Furthermore,
regardless of all profound examinations on the singularity area of the BH,
where extreme curvature effects exist is unknown and is an open problem in
physics, unless we develop a quantum gravity theory \cite{3}. In order to solve
the singularity problem various models of BHs are introduced. In recent years,
these non-singular BH solutions called regular BHs have attracted much attention,
particularly the non-linear electrodynamics model coupled to Einstein's gravity theory.
Firstly, Bardeen proposed that the regular BHs by incorporating magnetic charges follow
the weak energy condition \cite{4} . Numerous different work of Bardeen-like BHs are conducted to
remove singularities by considering nonlinear electrodynamics \cite{5}-\cite{10}.
Therefore, it has worth to examine their observational signatures \cite{11}.
Among the various alternatives to non-trivial topology in Einstein's equations,
those with topological defects of different forms are of particular interest.
Such space-times are also influenced by considerations in particle physics and
cosmology in addition to their basic geometric properties. Specifically, these defects known
as cosmic strings generally occur as a result of unexpected breaking of symmetry in the
theories of Yang-Mills \cite{12}. The existence of a cosmic string is expressed in the
non-triviality of the very first homotopy group of space-time, indicating that there are
closed curves accompanying the string which can never be deformed to a point continuously.
The space-time with cosmic strings and their singularity structures have been studied in
literature \cite{13}-\cite{16}.

The covariant conservation of the energy-momentum tensor is an important part in GR and
in many other modified theories, which leads to the conservation of certain physical
quantities through the Noether symmetry theorem.  Although, some modified theories have
suggested that it is possible to relax the conservation condition of energy-momentum tensor.
In $1972$, Rastall introduced a new gravity theory from these theories in which the standard
conservation law is questioned \cite{17}. The standard conservation law of energy-momentum tensor
is not satisfied in Rastall gravity theory \cite{18}. A non-minimal coupling of matter field
via geometry is taken into account in this gravitational framework that satisfies the following
relation $\nabla^{a}T_{ab}=\nabla_{b}R$.
As a strong point, this proposal has the fact that the standard energy-momentum tensor conservation
law is only checked in flat space time or in a flat space-time or in a
weak gravitational field limit. Numerous studies were made with this theory. Kumar and Ghosh \cite{19}
have studied the solution of rotating BH surrounded by perfect fluid in Rastall theory that rotating
Rastall BH generalize the Kerr-Newman BH solution. They have also analyzed the thermodynamics
(i.e., horizon area, entropy, rotational velocity and temperature)of rotating BH in the context of Rastall gravity.
Many researchers have discussed the Hawking temperature and corrected temperature by taking 
into account the generalized uncertainty principle (GUP) effects for various type of BHs in literature \cite{20}-\cite{29}.
By considering GUP effects, it is very feasible to evaluate the quantum corrected thermodynamics of BHs \cite{30}.
The GUP relation satisfies the following expression
\begin{equation*} 
\Delta x\Delta p \geq  \frac{1}{2}\Big(1+\beta(\Delta p)^2\Big).
\end{equation*}
here $\beta=\frac{\beta_{\circ}}{M^{2}_{p}}$, $M^{2}_{p}$ represents the Plank's mass 
and $\beta_{\circ}$ stands for dimensionless parameter. Li and Zu \cite{31} have analyzed 
the GUP effects by considering for spin-$0$ particles and
also investigated the corrected temperature.

The main objective of this paper is to investigate the RBHCS solution in the context of Newman-Janis algorithm and to derive the RBHCS Hawking temperature($T_{H}$) under the influences of rotation parameter as well as to
give a comparison of our new results with previous literature with the help of graphs.
Moreover, to compute the quantum corrected $T_{H}$ for RBHCS in Newman-Janis algorithm
accompanying GUP effects and analyze the stable condition of BH in the presence of quantum gravity effects.

In Sec. \textbf{II},
comprise RBHCS solution in the background of Newman-Janis algorithm and gives
the $T_{H}$ for the RBHCS in the Newman-Janis algorithm. Sec.\textbf{III} analyze the graphical representation of
$T_{H}$ with event horizon and describes the physical and stable condition of RBHCS under
the spin parameter effects. Section \textbf{IV} investigate the quantum
corrected $T_{H}$ for RBHCS in Newman-Janis algorithm.
Section \textbf{V} study the physical state of BH in the presence of quantum gravity effects. Section \textbf{VI} study the entropy for regular RBHCS with rotation parameter presence of gravity parameter effects. Section \textbf{VII} describes the graphical analysis of the logarithmic corrected entropy. At last, Sec. \textbf{VIII} conducts the final results of our work.
\section{Regular Black Holes with Cosmic Strings in Newman-Janis algorithm}
In the Newman-Janis algorithmic rule, which is obtained the new solution with only rotation parameters. We compute the RBCS solution in the background of Newman-Janis algorithmic rule by applying the approach \cite{Ra1}-\cite{Re5}. After calculating the RBCS solution in the background of rotation parameter, we examine the thermodynamic property of RBCS.
The RBHCS in $4$-dimensional space-time can be given as \cite{R1}
\begin{equation}
ds^{2}=-H(r)dt^2+H^{-1}(r)dr^2+r^2 d\theta^2+r^2 \sin^2\theta \zeta^2d\phi^2,\label{1}
\end{equation}
where the cosmic string parameter is $\zeta^2=1-4\mu$ and $\mu$ represents the mass per unit length of the string. The metric function $H(r)$
and mass function for RBHCS can be defined as
\begin{equation}
H(r)=1-2r^{-1}M(r), \nonumber
\end{equation}
with
\begin{equation}
M(r)=\frac{m_{\circ}}{\left[1+(\frac{r_{\circ}}{r})^{q}\right]^{\frac{p}{q}}},\nonumber
\end{equation}
 where $r_{\circ}$ and $m_{\circ}$ represents the length and mass of the string, respectively. We recover the
Bardeen as well as BH Hayward metric for the values ($q=2$, $p=3$ \& $q=p=3$), respectively \cite{R2}.
For the case $r_{\circ}<m_{\circ}$, we get two solutions of $M(r)=0$ as $r=r_{\pm}$. Here  $r_{-}$ represents
the inner horizon and $r_{+}\approx 2M(r)$ shows the outer horizon.

Now we introduce a new coordinate transformation in the devoted form
\begin{eqnarray}
du=dt-\frac{dr}{H(r)}.\label{A}
\end{eqnarray}
According to above transformation the metric (\ref{1}) gets the form
\begin{equation}
ds^{2}=-H(r)du^2-2dudr+r^2 d\theta^2+r^2 \zeta^2\sin^2\theta d\phi^2.
\end{equation}
The non-zero inverse elements of the metric can be defined as
\begin{equation}
g^{01}=g^{10}=-1,~~g^{11}=H(r),~~g^{22}=\frac{1}{r^2},~~g^{33}
=\frac{1}{r^2 \zeta^2\sin^2\theta}.\nonumber
\end{equation}
In the null tetrad frame the metric can be written as
\begin{eqnarray}
g^{\mu\nu}=-l^\nu n^\mu-l^\mu n^\nu+m^\mu \bar{m}^{\nu}+m^\nu \bar{m}^{\mu}.
\end{eqnarray}
The corresponding components of the null tetrad are given as
\begin{eqnarray}
l^{\mu}&=&\delta_{r}^{\mu},~~~n^{\nu}=\delta_{u}^{\mu}-\frac{H(r)}{2}\delta_{r}^{\mu},\nonumber\\
m^{\mu}&=&\frac{1}{r\sqrt{2}} \delta_{\theta}^{\mu}+\frac{i}{\sqrt{2}r
\zeta^2 \sin\theta}\delta_{\phi}^{\mu},\nonumber\\
\bar{m}^{\mu}&=&\frac{1}{r\sqrt{2}} \delta_{\theta}^{\mu}-\frac{i}
{\sqrt{2}r \zeta^2 \sin^2\theta}\delta_{\phi}^{\mu}.\nonumber
\end{eqnarray}
The null vectors of the null tetrad satisfy the following relations for
any point in the BH space-time $l_{\mu}l^{\mu}=n_{\mu}n^{\mu}=m_{\mu}m^{\mu}
=l_{\mu}m^{\mu}=m_{\mu}m^{\mu}=0$ and $l_{\mu}n^{\nu}=-m_{\mu}\bar{m}^{\mu}=1$.
In $(u, r)$ plane, we consider a coordinate transformation \cite{Rb2} as $u\rightarrow
u-ia\cos\theta$, $r\rightarrow r+ia\cos\theta$ and $\frac{2M}{r}\rightarrow \frac{M}{r}+\frac{M}{\bar{r}}=\frac{2Mr}{r^{2}+a^{2}cos^{2}\theta}$. Also, we perform the transformation
$H(r)\rightarrow \tilde{H}(r, a, \theta)$ and $\sigma^2=a^2 \cos^2\theta+r^2$.
In space $(u, r)$, the vectors of null obtain the following form
\begin{eqnarray}
l^{x\mu}&=&\delta_{r}^{\mu},~~~n^{\nu}=\delta_{u}^{\mu}-\frac{1}{2}
\tilde{H}(r) \delta_{r}^{\mu},\nonumber\\m^{\mu}&=&\frac{1}{\sqrt{2}r}
\left(\delta_{\theta}^{\mu}+ia \zeta^2 \sin\theta\left(\delta_{u}^{\mu}
-\delta_{r}^{\mu}\right)+\frac{i}{\zeta^2 \sin\theta}\delta_{\phi}^{\mu}
\right),\nonumber\\\bar{m}^{\mu}&=&\frac{1}{r\sqrt{2}}\left(\delta_{\theta}
^{\mu}-ia \zeta^2 \left(\delta_{u}^{\mu} -\delta_{r}^{\mu}\right)\sin\theta
-\frac{i}{\zeta^2 \sin\theta}\delta_{\phi}^{\mu}\right).\nonumber
\end{eqnarray}
From the definition of the null tetrad the inverse metric $g^{\mu\nu}$ in
the Eddington-Finkelstein coordinate is given by
\begin{eqnarray}
g^{00}&=&\frac{a^2\zeta^2 sin^2\theta}{\sigma^2},~~~g^{01}=g^{10}=-1-\frac{a^2
\zeta^2 sin^2\theta}{\sigma^2},~~~g^{11}=\tilde{H}(r)+\frac{a^2 \zeta^2
\sin^2\theta}{\sigma^2},~~~g^{22}=\frac{1}{\sigma^2},\nonumber\\
g^{33}&=&\frac{1}{\sigma^2 \zeta^2 \sin^2\theta},~~~g^{03}=g^{30}=\frac{a}
{\sigma^2},~~~g^{13}=g^{31}=-\frac{a}{\sigma^2},\nonumber
\end{eqnarray}
and lower indices components of the metric are
\begin{eqnarray}
g_{00}&=&-\tilde{H},~~~g_{01}=g_{10}=-1,~~~g_{11}=0,~~~
g_{22}=\sigma^2,~~~g_{13}=g_{31}=-\frac{a}{\sigma^2}\nonumber\\
g_{33}&=&\zeta^2 \sin^2\theta\left(a^2(\tilde{H}(r)-2)\zeta^2
\sin^2\theta+\sigma^2\right),~~~g_{03}=g_{30}=a \zeta^2 \sin^2\theta,\nonumber
\end{eqnarray}
here
\begin{equation}
\tilde{H}(r, \theta)=1-\frac{2Mr}{\sigma^2}.
\end{equation}
Finally, we perform the coordinate transformation from Eddington-Finkelstein
to Boyer-Lindquist coordinates as
\begin{equation}
du=dt+\Upsilon(r)dr,~~~d\phi=d\phi+\xi(r)dr,\label{a1}
\end{equation}
where the $\Upsilon(r)$ and $\xi(r)$ functions are introduced to extract
the $g_{r\phi}$ and $g_{tr}$ components as
\begin{equation}
\Upsilon(r)=\frac{r^2+a^2}{r^{2}\tilde{H}(r) +a^2},~~~\xi(r)=-\frac{a}
{r^{2}\tilde{H}(r)+a^2}.
\end{equation}
Ultimately, having the above expressions of $\Upsilon(r)$ and $\xi(r)$, we get
the RBHCS solution in the background of Newman-Janis algorithm as
\begin{equation}
ds^2=-\tilde{H}(r)dt^2+\frac{\sigma^2}{\Delta}dr^2+2a\left(\tilde{H}(r)-1\right)\zeta^2 sin^2\theta dt d\phi+ \sigma^2 d\theta^2
+\zeta^2 sin^2\theta\left(\sigma^2+a^2\zeta^2 \left(2-\tilde{H}(r)\right)sin^2\theta\right)d\phi^2. \label{cc}
\end{equation}
The above equation implies
\begin{equation}
ds^{2}=-\left(1-\frac{2Mr}{\sigma^2}\right)dt^2-2a\zeta^2\left(\frac{2Mr}
{\sigma^2}\right)\sin^2\theta dt d\phi+\frac{\sigma^2}{\Delta}dr^2
+\sigma^2d\theta^2+\zeta^2\sin^2\theta\left(\sigma^2+\frac{2Mr}{\sigma^2}
a^2\zeta^2\sin^2\theta\right)d\phi^2, \label{01}
\end{equation}
where $\Delta=r^2+a^2-2Mr$.
The angular velocity of horizon ($\Omega$) and angular momentum ($J$) are given as 
\begin{equation*}
\Omega=\frac{2Mar}{\sigma^{4}+2Mar\zeta^2},~~~~
J=\frac{Mar(r^{2}+a^2)}{\sigma^{4}+2Mar\zeta^2}.
\end{equation*}
It is worth to mention here that, by utilizing
the Newman-Janis algorithmic rule to obtain the rotating solution and also discussed that the regular solution \cite{NJ, NJ1} of space-time.
The $T_{H}$ can be conducted from the following formula
\begin{equation}
T_{H} = \frac{\tilde{H}'(r_+)}{4\pi}.
\end{equation}
The corresponding $T_{H}$ for RBHCS in the Newman-Janis algorithm at $r=r_+$ is given as
\begin{equation}
T_{H} = \frac{m_{\circ}\left[1+\left(\frac{r_\circ}{r_+}\right)^q\right]^{-\frac{p+q}{q}}
\left[1-(p-1)\left(\frac{r_\circ}{r_+}\right)^q\right]}{2\pi\left(r^2_++a^2\right)}.\label{22}
\end{equation}
This $T_{H}$ depends upon the mass $m_{\circ}$, length $r_{\circ}$, BH radius $r_+$ and rotation parameter $a$.
We get the $T_{H}$ for RBHCS \cite{TH} in the absence of rotation parameter (i.e., $a=0$).
It has worth to mention here that, by considering $m_\circ=m, r_{\circ}=0=a$ at
$r_+\thickapprox2m$, we recover the Schwarzschild Hawking temperature $T_{Sch}=1/8\pi m$.
The RBHCS in the context of Newman-Janis algorithm are colder than the Schwarzschild BH. In the next section,
we graphically investigate this phenomenon.
\section{$T_{H}$ Analysis}
We examine the physical
significance of temperature $T_{H}$ via horizon $r_{+}$ in the presence/absence of rotation parameter
and analyze the physical and stable condition of the corresponding RBHCS.
Based on the Hawking phenomenon, as the temperature rises and more radiation is
released, the BH radius decreases. The stability of BH is defined by this physical phenomenon.

\textbf{Fig. 1}: (i) shows the presentation of $T_{H}$ for fixed $m_{\circ}=50$, $r_{\circ}=0.001$,
$p=3, q=2$ and rotation parameter $a$ varying values in the range $0\leq r_{+}\leq5$. Firstly,
the $T_{H}$ depicts constant behavior but after getting a maximum height the
$T_{H}$ gradually reductions with the increasing $r_+$ and obtain an asymptotically
flat sate as $r_{+}\rightarrow\infty$. This behavior comprise the stable state of BH.
Moreover, we also observe the BH remnant (maximum $T_{H}$ at non-zero $r_{+}$).

(ii) represents the behavior of $T_{H}$ via $ r_{+}$ for varying values
of mass $m$ in the absence of spin parameter and length (i.e., $a=0=r_{\circ}$) in the range $0\leq r_{+}\leq5$.
The decreasing $T_{H}$ with raising horizon shows the physical state of BH.
In the previous section, we have analyzed that the RBHCS in the context of Newman-Janis algorithm.
From both plots, one can observe that the
$T_{H}$ in plot (ii) is higher than in plot (i), when we consider $m_\circ=m, r_{\circ}=0=a$ in Eq. (\ref{22}).
\begin{center}
\includegraphics[width=7cm]{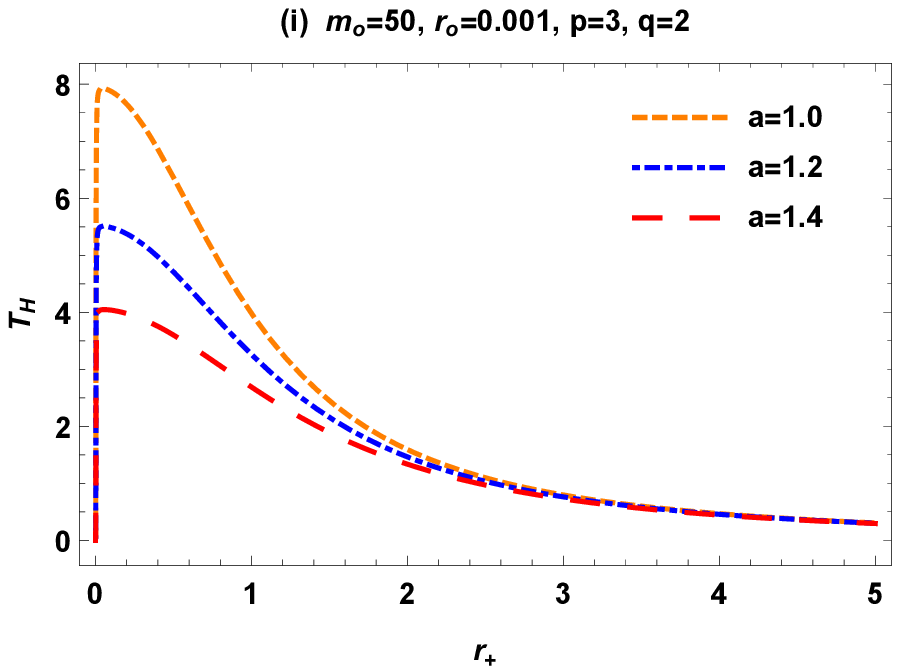}\includegraphics[width=7cm]{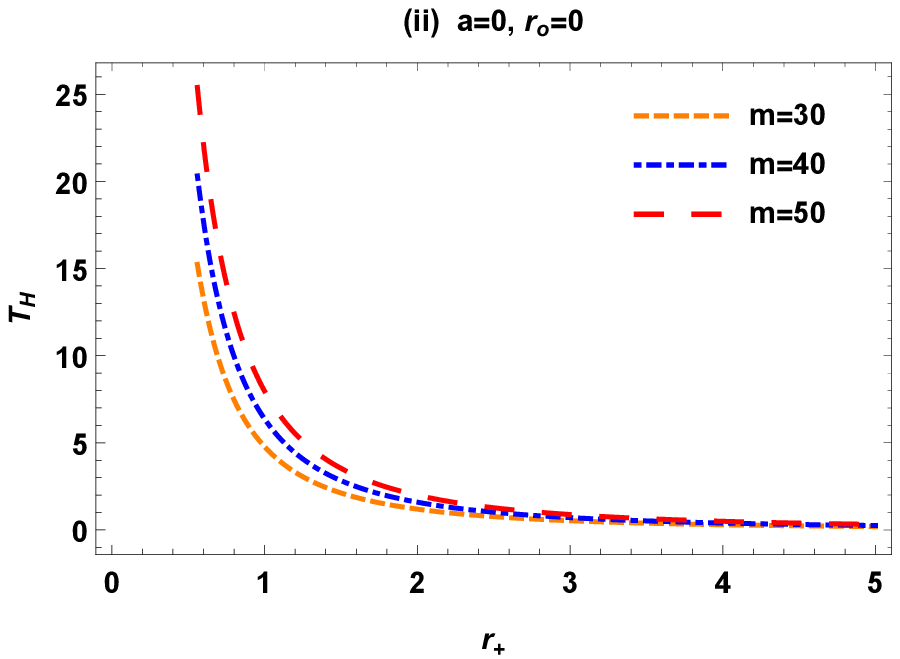}\\
{Figure 1: $T_{H}$ versus $r_{+}$.}
\end{center}
\section{Quantum Corrected Temperature for Regular Black Holes with Cosmic Strings in Newman-Janis Algorithm}
In this section, we investigate the effects of quantum gravity in RBHCS with spin parameter solution.
We analyze the corrected temperature for the metric (\ref{cc}) by incorporating GUP effects.
The modified equation \cite{29a} of motion for spin-$1$ particles can be given as
\begin{eqnarray}
&&\partial_{\mu}(\sqrt{-g}\chi^{\nu\mu})+\frac{m^2}{\hbar^2}
\sqrt{-g}\chi^{\nu}+\frac{i}{\hbar}\sqrt{-g}A_{\mu}\chi^{\nu\mu}+\frac{i}
{\hbar}\sqrt{-g}eF^{\nu\mu}\chi_{\mu}+\hbar^{2}\beta\partial_{0}\partial_{0}
\partial_{0}(g^{00}\sqrt{-g}\chi^{0\nu})\nonumber\\
&&-\hbar^{2}\beta\partial_{i}\partial_{i}\partial_{i}(g^{ii}\sqrt{-g}\chi^{i\nu})=0,\label{A1}
\end{eqnarray}
here $\chi^{\nu\mu}$, $g$ and $m$ show the
anti-symmetric tensor, the determinant of coefficient matrix and the particle mass. The $\chi_{\nu\mu}$ can be defined as
\begin{eqnarray}
\chi_{\nu\mu}&=&(1-\hbar^2\beta\partial_{\nu}^2)\partial_{\nu}\chi_{\mu}-
(1-\hbar^2\beta\partial_{\mu}^2)\partial_{\mu}\chi_{\nu}+(1-\hbar^2\beta\partial_{\nu}^2)\frac{i}{\hbar}eA_{\nu}\chi_{\mu}-(1-\hbar^2\beta\partial_{\nu}^2)\frac{i}{\hbar}eA_{\mu}\chi_{\nu},~~\text{and}~~\nonumber\\
F_{\nu\mu}&=&\nabla_{\nu} A_{\mu}-\nabla_{\mu} A_{\nu},\nonumber
\end{eqnarray}
where $~e~$,  $\nabla_{\mu}$,  $\beta$, and $A_{\mu}$ are the particle charge, covariant derivative, GUP parameter and
RBHCS vector potential respectively.

In corrected $T_H$ for spin-$1$ particles,
we rewrite the metric Eq. (\ref{cc}) in the proceeding form
\begin{eqnarray}
ds^{2}&=&-\tilde{A}dt^{2}+\tilde{B}dr^{2}+\tilde{C}d\theta^{2}
+\tilde{D} d\phi^{2}+2\tilde{E}dt d\phi,\label{aa}
\end{eqnarray}
where
\begin{equation}
\tilde{A}=\tilde{H}(r),~~\tilde{B}=\frac{\sigma^2}{\Delta},
~~\tilde{C}=\sigma^2,~~\tilde{D}=\zeta^2 \sin^2\theta\left(\sigma^2
+a^2\zeta^2 \left(2-\tilde{H}(r)\right)\sin^2\theta\right),
~~\tilde{E}=a\zeta^2\left(\tilde{H}(r)-1\right) \sin^2\theta.\nonumber
\end{equation}
After using the metric Eq. (\ref{aa}), the components of ${\chi}^{\mu}$ and
${\chi}^{\nu\mu}$ can be computed as
\begin{eqnarray}
\chi^{0}&=&\frac{\tilde{-D}\chi_{0}+\tilde{E}\chi_{3}}{\tilde{A}\tilde{D}
+\tilde{E}^2},~~~~~\chi^{1}=\frac{1}{\tilde{B}}\chi_{1},~~~\chi^{2}
=\frac{1}{\tilde{C}}\chi_{2},~~~\chi^{3}=\frac{\tilde{E}\chi_{0}
+\tilde{A}\chi_{3}}{\tilde{A}\tilde{D}+\tilde{E}^2},~~~~~\chi^{01}
=\frac{\tilde{-D}\chi_{01}+\tilde{E}\chi_{13}}{(\tilde{E}^2+\tilde{A}\tilde{D}
)\tilde{B}},\nonumber\\~~~\chi^{02}&=&\frac{\tilde{-D}\chi_{02}}{
(\tilde{E}^2+\tilde{A}\tilde{D})\tilde{C}},~~~~\chi^{03}=\frac{(\tilde{A}^2-\tilde{A}\tilde{D})\chi_{03}}{(\tilde{A}\tilde{D}+\tilde{E}^2)^2},~~\chi^{12}=\frac{1}{\tilde{B}
\tilde{C}}\chi_{12},~~\nonumber\\
\chi^{13}&=&\frac{1}{\tilde{B}(\tilde{A}
\tilde{D}+\tilde{E}^2)}\chi_{13},~~\chi^{23}=\frac{\tilde{A}\chi_{23}+\tilde{E}\chi_{02}}
{(\tilde{A}\tilde{D}+\tilde{E}^2)\tilde{C}}.\nonumber
\end{eqnarray}
The approximation of WKB is defined \cite{29a} as
\begin{equation}
\chi_{\nu}=s_{\nu}\exp\left[\frac{i}{\hbar}I_{0}(t,r,\phi, \theta)+
\Sigma \hbar^{n}I_{n}(t,r, \phi, \theta)\right].\label{ab}
\end{equation}
where $I_{0}$ and $I_n$ represent to particle action, for $n= 1, 2, 3, . . .$ and neglecting the higher order because $\hbar$ is a Planck constant, and the higher power of $\hbar n$ terms is very small in \cite{R2a, R2b, R2c, R2d, R2e}.
After substituting the Eqs. (\ref{ab}) into Eq. (\ref{A1}),
we get a set of field equations as
\begin{eqnarray}
&&\frac{\tilde{D}}{(\tilde{E}^2+\tilde{A}\tilde{D})\tilde{B}}
\left[s_{1}(\partial_{0}I_{0})(\partial_{1}I_{0})+\beta s_{1}
(\partial_{0}I_{0})^{3}(\partial_{1}I_{0})-s_{0}(\partial_{1}I_{0})^{2}
-\beta s_{0}(\partial_{1}I_{0})^4+s_{1}eA_{0}(\partial_{1}I_{0})\right.\nonumber\\
&+&\left.s_{1}\beta eA_{0}(\partial_{0}I_{0})^{2}(\partial_{1}I_{0})\right]
-\frac{\tilde{E}}{\tilde{B}(\tilde{A}\tilde{D}+\tilde{E}^2)}
\left[s_{3}(\partial_{1}I_{0})^2+\beta s_{3}(\partial_{1}I_{0})^4
-s_{1}(\partial_{1}I_{0})(\partial_{3}I_{0})-\beta s_{1}
(\partial_{1}I_{0})(\partial_{3}I_{0})^2\right]\nonumber\\
&+&\frac{\tilde{D}}{\tilde{C}(\tilde{A}\tilde{D}+\tilde{E}^2)}
\left[s_{2}(\partial_{0}I_{0})(\partial_{2}I_{0})+\beta s_{2}
(\partial_{0}I_{0})^3(\partial_{2}I_{0})-s_{0}(\partial_{2}I_{0})^2
-\beta s_{0}(\partial_{2}I_{0})^4+s_{2}eA_{0}(\partial_{2}I_{0})\right.\nonumber\\
&+&\left.s_{2}eA_{0}\beta(\partial_{0}I_{0})^{2}(\partial_{1}I_{0})\right]
+\frac{\tilde{A}\tilde{D}}{(\tilde{A}\tilde{D}+\tilde{E}^2)^2}
\left[s_{3}(\partial_{0}I_{0})(\partial_{3}I_{0})+\beta s_{3}
(\partial_{0}I_{0})^{3}(\partial_{3}I_{0})-s_{0}(\partial_{3}I_{0})^{2}
-\beta s_{0}(\partial_{3}I_{0})^4\right.\nonumber\\
&+&\left. s_{3}eA_{0}(\partial_{3}I_{0})+s_{3}eA_{0}(\partial_{0}I_{0})^{2}
(\partial_{3}I_{0})\right]-m^2\frac{\tilde{D s_{0}}-\tilde{E s_{3}}}
{(\tilde{A}\tilde{D}+\tilde{E}^2)}=0,\label{j1}\\
&-&\frac{\tilde{D}}{\tilde{B}(\tilde{A}\tilde{D}+\tilde{E}^2)}
\left[s_{1}(\partial_{0}I_{0})^2+\beta s_{1}(\partial_{0}I_{0})^4-s_{0}
(\partial_{0}I_{0})(\partial_{1}I_{0})-\beta s_{0}(\partial_{0}I_{0})(\partial_{1}I_{0})^{3}
+s_{1}eA_{0}(\partial_{0}I_{0})
+\beta s_{1}eA_{0}(\partial_{0}I_{0})^3\right]\nonumber
\end{eqnarray}
\begin{eqnarray}
&+&\frac{\tilde{E}}{\tilde{B}
(\tilde{A}\tilde{D}+\tilde{E}^2)}\left[s_{3}(\partial_{0}I_{0})(\partial_{1}I_{0})
+\beta s_{3}(\partial_{0}I_{0})
(\partial_{1}I_{0})^3-s_{1}(\partial_{0}I_{0})(\partial_{3}I_{0})
-\beta s_{1}(\partial_{0}I_{0})(\partial_{3}I_{0})^{3}\right]
+\frac{1}{\tilde{B}\tilde{C}}\left[s_{2}(\partial_{1}I_{0})(\partial_{2}I_{0})\right.\nonumber\\
&+&\left.\beta s_{2}
(\partial_{1}I_{0})(\partial_{2}I_{0})^3-s_{1}(\partial_{2}I_{0})^{2}
-\beta s_{1}(\partial_{2}I_{0})^{4}\right]+\frac{1}{\tilde{B}(\tilde{A}\tilde{D}
+\tilde{E}^2)}\left[s_{3}(\partial_{1}I_{0})(\partial_{3}I_{0})
+\beta s_{3}(\partial_{1}I_{0})(\partial_{3}I_{0})^3-s_{1}
(\partial_{3}I_{0})^2\right.\nonumber\\
&-&\left.\beta s_{1} (\partial_{3}I_{0})^{4}\right]-\frac{m^2 s_{1}}{\tilde{B}}
+\frac{eA_{0}\tilde{D}}{\tilde{B}(\tilde{A}\tilde{D}+\tilde{E}^2)}
\left[s_{1}(\partial_{0}I_{0})+\beta s_{1}(\partial_{0}I_{0})^3
-s_{0}(\partial_{1}I_{0})-\beta s_{0}(\partial_{1}I_{0})^3
+ eA_{0}s_{1}\right.\nonumber\\
&+&\left.\beta s_{1}eA_{0}(\partial_{0}I_{0})^{2})\right]
+\frac{eA_{0}\tilde{E}}{\tilde{B}(\tilde{A}\tilde{D}+\tilde{E}^2)}
\left[s_{3}(\partial_{1}I_{0})+\beta s_{3}(\partial_{1}I_{0})^3
-s_{1}(\partial_{3}I_{0})-\beta s_{1}(\partial_{1}I_{0})^3\right]=0,\\
&+&\frac{\tilde{D}}{\tilde{C}(\tilde{A}\tilde{D}+\tilde{E}^2)}
\left[s_{2}(\partial_{0}I_{0})^2+\beta s_{2}(\partial_{0}I_{0})^{4}
-s_{0}(\partial_{0}I_{0})(\partial_{2}I_{0})-\beta s_{0}(\partial_{0}I_{0})(\partial_{2}I_{0})^3
+s_{2}eA_{0}(\partial_{0}I_{0})+\beta s_{2}eA_{0}(\partial_{0}I_{0})^{3}\right]\nonumber\\
&+&\frac{1}{\tilde{B}\tilde{C}}\left[s_{2}(\partial_{1}I_{0})^2+\beta s_{2}
(\partial_{1}I_{0})^{4}-s_{1}(\partial_{1}I_{0})(\partial_{2}I_{0})
-\beta s_{1}(\partial_{1}I_{0})(\partial_{2}I_{0})^3\right]-\frac{\tilde{E}}
{\tilde{C}(\tilde{A}\tilde{D}+\tilde{E}^2)}\Big[s_{2}(\partial_{0}I_{0})
(\partial_{3}I_{0})\nonumber\\
&+&\left.\beta s_{2}(\partial_{0}I_{0})^{3}(\partial_{3}I_{0})-s_{0}
(\partial_{0}I_{0})(\partial_{3}I_{0})-\beta s_{0}(\partial_{0}I_{0})^3
(\partial_{3}I_{0})+s_{2}eA_{0}(\partial_{3}I_{0})+\beta s_{2}eA_{0}
(\partial_{3}I_{0})^{3}\right]\nonumber\\
&+&\frac{\tilde{A}}{\tilde{C}(\tilde{A}\tilde{D}+\tilde{E}^2)}\left[s_{3}
(\partial_{2}I_{0})(\partial_{3}I_{0})+\beta s_{3}
(\partial_{2}I_{0})^{3}(\partial_{3}I_{0})-s_{2}(\partial_{3}I_{0})^2
-\beta s_{2}(\partial_{3}I_{0})^4\right]+\frac{eA_{0}\tilde{D}}{\tilde{C}
(\tilde{A}\tilde{D}+\tilde{E}^2)}\Big[s_{2}(\partial_{0}I_{0})\nonumber\\
&+&\left.\beta s_{2}(\partial_{0}I_{0})^3-s_{0}(\partial_{2}I_{0})
-\beta s_{0}(\partial_{2}I_{0})^3+s_{2}eA_{0}+s_{2}\beta eA_{0}
(\partial_{0}I_{0})^2\right]-\frac{m^2 s_{2}}{\tilde{C}}=0,\\
&+&\frac{(\tilde{A}\tilde{D})-\tilde{A}^2}{(\tilde{A}\tilde{D}
+\tilde{E}^2)^2}\left[s_{3}(\partial_{0}I_{0})^2+\beta s_{3}
(\partial_{0}I_{0})^4-s_{0}(\partial_{0}I_{0})(\partial_{3}I_{0})
-\beta s_{0}(\partial_{0}I_{0})(\partial_{3}I_{0})^{3}+{eA_{0}s_3}
(\partial_{0}I_{0})+\beta s_{3}eA_{0}(\partial_{0}I_{0})^{3}\right]\nonumber\\
&-&\frac{\tilde{D}}{\tilde{C}(\tilde{A}\tilde{D}+\tilde{E}^2)}
\left[s_{3}(\partial_{1}I_{0})^2+\beta s_{3}(\partial_{1}I_{0})^{4}
-s_{1}(\partial_{1}I_{0})(\partial_{3}I_{0})-\beta s_{1}(\partial_{1}
I_{0})(\partial_{3}I_{0})^3\right]-\frac{\tilde{E}}{\tilde{C}(\tilde{A}
\tilde{D}+\tilde{E}^2)}\Big[s_{2}(\partial_{0}I_{0})(\partial_{2}I_{0})\nonumber\\
&+&\left.\beta s_{2}(\partial_{0}I_{0})^3(\partial_{2}I_{0})-s_{0}
(\partial_{2}I_{0})^{2}+\beta s_{0}(\partial_{2}I_{0})^4+{eA_{0}s_2}
(\partial_{2}I_{0})+\beta s_{2}eA_{0}\partial_{0}I_{0})^{2}(\partial_{2}
I_{0})\right]-\frac{eA_{0}\tilde{A}}{\tilde{C}(\tilde{A}\tilde{D}+\tilde{E}^2)}
\left[s_{3}(\partial_{2}I_{0})^2\right.\nonumber\\
&+&\left.\beta s_{3}(\partial_{2}I_{0})^4-s_{2}(\partial_{2}I_{0})(\partial_{3}I_{0})
-\beta s_{2}(\partial_{0}I_{0})(\partial_{3}I_{0})^{3}\right]
-\frac{m^2 (\tilde{E}s_{0}-\tilde{A}s_{3}}{(\tilde{A}\tilde{D}+\tilde{E}^2)}
+\frac{eA_{0}(\tilde{A}\tilde{D})-\tilde{A}^2}{(\tilde{A}\tilde{D}+\tilde{E}^2)^2}
\left[s_{3}(\partial_{0}I_{0})+\beta s_{3}
(\partial_{0}I_{0})^3\right.\nonumber\\
&-&\left.s_{0}(\partial_{3}I_{0})-\beta s_{0}(\partial_{3}I_{0})^3+s_{3}seA_{0}
+s_{3}\beta eA_{0}(\partial_{0}I_{0})^2\right]=0,\label{j2}
\end{eqnarray}
Using separation of variables technique, we can choose
\begin{equation}
I_{0}=-(\hat{E}-J\omega)t+W(r)+j\phi+\nu(\theta),\label{ff}
\end{equation}
where $\hat{E}$ and $J$  represents the energy and angular
momentum of the particles, respectively, corresponding to angle $\phi$.

After substituting Eq. (\ref{ff}) into Eqs. (\ref{j1})-(\ref{j2}), we attain a matrix in the form
\begin{equation*}
K(s_{0},s_{1},s_{2},s_{3})^{T}=0,
\end{equation*}
which implies a $4\times4$ matrix labeled as "$K$", whose components
are given as follows:
\begin{eqnarray}
K_{00}&=&\frac{-\tilde{D}}{\tilde{B}(\tilde{A}\tilde{D}+\tilde{E}^2)}
\left[W_{1}^2+\beta W_{1}^4\right]-\frac{\tilde{D}}{(\tilde{A}\tilde{D}+\tilde{E}^2)\tilde{C}}\Big[j^2+\beta j^4\Big]
-\frac{\tilde{A}\tilde{D}}{(\tilde{A}\tilde{D}+\tilde{E}^2)^2}
\Big[\nu_{1}^2+\beta \nu_{1}^4\Big]-\frac{m^2 \tilde{D}}{(\tilde{A}
\tilde{D}+\tilde{E}^2)},\nonumber\\
K_{01}&=&\frac{-\tilde{D}}{\tilde{B}(\tilde{A}\tilde{D}+\tilde{E}^2)}
\Big[(E+\beta E^3+eA_{0}+\beta eA_{0}E^2\Big]W_{1}
+\frac{\tilde{E}}{\tilde{B}(\tilde{A}\tilde{D}+\tilde{E}^2)}+\Big[\nu_{1}+
\beta \nu_{1}^3\Big],\nonumber\\
K_{02}&=&\frac{-\tilde{D}}{\tilde{C}(\tilde{A}\tilde{D}+\tilde{E}^2)}
\Big[E+\beta \hat{E}^3-eA_{0}-\beta eA_{0}E^2
\Big]j,\nonumber\\
K_{03}&=&\frac{-\tilde{E}}{\tilde{B}(\tilde{A}\tilde{D}+\tilde{E}^2)}
\Big[W_{1}^2+\beta W_{1}^4\Big]- \frac{\tilde{A}\tilde{D}}{\tilde{C}
(\tilde{A}\tilde{D}+\tilde{E}^2)^2}\Big[\beta E^3
-\beta eA_{0}E^2+E-eA_{0}\Big]\nu_{1}\nonumber\\
&+&\frac{m^2\tilde{E}}{(\tilde{A}\tilde{D}+\tilde{E}^2)^2},~~~~~
K_{12}=\frac{1}{\tilde{B}\tilde{C}}\Big[W_{1}+\beta W_{1}^3\Big]j
=K_{21},\nonumber
\end{eqnarray}
\begin{eqnarray}
K_{11}&=&\frac{-\tilde{D}}{\tilde{B}(\tilde{A}\tilde{D}+\tilde{E}^2)}
\Big[\beta E^4-\beta eA_{0}EW_{1}^2+E^2-eA_{0}E\Big]
+\frac{\tilde{E}}{(\tilde{A}\tilde{D}+\tilde{E}^2)\tilde{B}}\nonumber\\
&+&\Big[\nu_{1}+\beta \nu_{1}^3\Big]E-\frac{1}{\tilde{B}
\tilde{C}}\Big[j^2+\beta j^4\Big]-\frac{1}{(\tilde{A}\tilde{D}
+\tilde{E}^2)\tilde{B}}\Big[\nu_{1}+\beta \nu_{1}^3\Big]-\frac{m^2}{\tilde{B}}
-\frac{eA_{0}\tilde{D}}{(\tilde{A}\tilde{D}+\tilde{E}^2)\tilde{B}}
\Big[E\nonumber\\
&+&\beta E^3-eA_{0}-\beta eA_{0}E^2\Big]
+\frac{eA_{0}\tilde{E}}{\tilde{B}(\tilde{A}\tilde{D}+\tilde{E}^2)}\Big[\nu_{1}+
\beta \nu_{1}^3\Big],\nonumber\\
K_{13}&=&\frac{-\tilde{E}}{\tilde{B}(\tilde{A}\tilde{D}+\tilde{E}^2)}\Big[W_{1}
+\beta W_{1}^3\Big]E+\frac{1}{\tilde{B}(\tilde{A}\tilde{D}
+\tilde{E}^2)^2}\Big[W_{1}+\beta W_{1}^3\Big]\nu_{1}+\frac{\tilde{E}eA_{0}}
{\tilde{B}(\tilde{A}\tilde{D}+\tilde{E}^2)}\Big[W_{1}+\beta W_{1}^3\Big],\nonumber\\
K_{20}&=&\frac{\tilde{D}}{\tilde{C}(\tilde{A}\tilde{D}+\tilde{E}^2)}
\Big[Ej+\beta Ej^3\Big]+\frac{\tilde{E}}{\tilde{C}
(\tilde{A}\tilde{D}+\tilde{E}^2)}\Big[E+\beta E^3\nu_{1}^2\Big]
-\frac{\tilde{D}eA_{0}}{\tilde{C}(\tilde{A}\tilde{D}+\tilde{E}^2)}\Big[j+\beta j^3\Big],\nonumber\\
K_{22}&=&\frac{\tilde{D}}{\tilde{C}(\tilde{E}^2+\tilde{A}\tilde{D})}\Big[+\beta E^4-\beta eA_{0}E+E^2
-eA_{0}E\Big]
-\frac{1}{\tilde{B}\tilde{C}}+\frac{\tilde{E}}{\tilde{C}(\tilde{E}^2+\tilde{A}\tilde{D})}\Big[\beta E^3\nonumber\\
&+&-\beta eA_{0}E^2-eA_{0}+E\Big]\nu_{1}
-\frac{\tilde{A}}{\tilde{C}(\tilde{E}^2+\tilde{A}\tilde{D})}\Big[\nu_{1}^2+
\beta \nu_{1}^4\Big]-\frac{m^2}{\tilde{C}}-\frac{eA_{0}\tilde{D}}{\tilde{C}
(\tilde{A}\tilde{D}+\tilde{E}^2)}
\Big[E\nonumber\\
&+&\beta E^3-eA_{0}-\beta eA_{0}E^2\Big],\nonumber~~~~~~~
K_{23}=\frac{\tilde{A}}{\tilde{C}(\tilde{A}\tilde{D}+\tilde{E}^2)}
\Big[j+\beta j^3\Big]\nu_{1},\nonumber\\
K_{30}&=&\frac{(\tilde{A}\tilde{D}-\tilde{A^2})}{(\tilde{A}\tilde{D}
+\tilde{E}^2)^2}\Big[\nu_{1}+\beta \nu_{1}^3\Big]E
+\frac{\tilde{E}}{\tilde{C}(\tilde{A}\tilde{D}+\tilde{E}^2)}
\Big[j^2+\beta j^4\Big]-\frac{m^2\tilde{E}}{(\tilde{A}\tilde{D}
+\tilde{E}^2)}-\frac{eA_{0}(\tilde{A}\tilde{D}-\tilde{A^2})}{(\tilde{A}\tilde{D}
+\tilde{E}^2)^2}\Big[\nu_{1}+\beta \nu_{1}^3\Big],\nonumber\\
K_{31}&=&\frac{1}{\tilde{B}(\tilde{A}\tilde{D}+\tilde{E}^2)}
\Big[\nu_{1}+\beta \nu_{1}^3\Big]W_{1},~~~~~K_{32}=\frac{\tilde{E}}
{\tilde{C}(\tilde{E}^2+\tilde{A}\tilde{D})}\Big[j+\beta j^3\Big]
E+\frac{\tilde{A}}{\tilde{C}(\tilde{E}^2+\tilde{A}\tilde{D})}
\Big[\nu_{1}+\beta \nu_{1}^3\Big]J,\nonumber\\
K_{33}&=&\frac{(\tilde{A}\tilde{D}-\tilde{A^2})}{(\tilde{A}\tilde{D}
+\tilde{E}^2)}\Big[E^2+\beta E^4-eA_{0}E
-\beta eA_{0}E^3\Big]-\frac{1}{\tilde{B}(\tilde{E}^2+\tilde{A}\tilde{D}
)}\Big[W_{1}^2+\beta W_{1}^4\Big]\nonumber\\
&-&\frac{\tilde{A}}{(\tilde{E}^2+\tilde{A}\tilde{D})\tilde{C}}
\Big[j^2+\beta j^4\Big]-\frac{m^2 \tilde{A}}{(\tilde{A}\tilde{D}+\tilde{E}^2)}
-\frac{eA_{0}(\tilde{A}\tilde{D}-\tilde{A^2})}{(\tilde{A}\tilde{D}
+\tilde{E}^2)}\Big[E+\beta E^3-eA_{0}E^2\Big],\nonumber
\end{eqnarray}
where $E=\hat{E}-j\omega$, $j=\partial_{\phi}I_{0}$, $W_{1}=\partial_{r}{I_{0}}$ and $\nu_{1}=\partial_{\theta}{I_{0}}$.
The solution for non-trivial then $|\textbf{K}|=0$, so we get
\begin{eqnarray}\label{a1}
ImW^{\pm}&=&\pm \int\sqrt{\frac{(E-eA_{0})^{2}
+X_{1}\Big[1+\beta\frac{X_{2}}{X_{1}}\Big]}{\frac{(\tilde{A}\tilde{D}+\tilde{E}^2)}
{\tilde{B}\tilde{D}}}}dr,\nonumber\\
&=&\pm i\pi\frac{(E-eA_{0})}{2\kappa(r_{+})}\Big[1+\beta\wp\Big],
\end{eqnarray}
with
\begin{eqnarray}
X_{1}&=&\Big[E
-eA_{0}\Big]\nu_{1}\frac{\tilde{B}\tilde{E}}{(\tilde{A}\tilde{D}+\tilde{E}^2)}+\frac{\tilde{A}\tilde{B}}{(\tilde{A}\tilde{D}+\tilde{E}^2)}\nu_{1}^2
-\tilde{B}m^2,\nonumber\\
X_{2}&=&\frac{\tilde{B}\tilde{D}}{(\tilde{A}\tilde{D}+\tilde{E}^2)}\Big[E^4+(eA_{0})^{2}E^{2}-2eA_{0}E^{3}\Big]
+\frac{\tilde{B}\tilde{E}}{\tilde{C}(\tilde{A}\tilde{D}+\tilde{E}^2)}\Big
[E^3-eA_{0}E^2\Big]\nu_{1}\nonumber\\
&-&\frac{\tilde{A}\tilde{B}}{(\tilde{A}\tilde{D}+\tilde{E}^2)}\nu_{1}^4-W_{1}^4.\nonumber
\end{eqnarray}
and 
\begin{equation*}
\wp=6m^{2}+\frac{6j^{2}}{r^{2}_{+}}
\end{equation*}
The charged particles tunneling probability can be computed as
\begin{equation}
\Gamma=\frac{\Gamma_{\textmd{emission}}}{\Gamma_{\textmd{absorption}}}=
\exp\Big[{-4\pi}\frac{(E-eA_{0})}
{2\kappa(r_{+})}\Big]\Big[1+\beta\wp\Big].
\end{equation}
The BH $\kappa$ (surface gravity) can be computed as
\begin{equation*}
\bar{\kappa}=\frac{m_{\circ}\left[1+\left(\frac{r_\circ}{r_+}\right)^q\right]^{-\frac{p+q}{q}}
\left[1-(p-1)\left(\frac{r_\circ}{r_+}\right)^q\right]}{\left(r^2_++a^2\right)}.
\end{equation*}
By considering Boltzmann factor for Hawking temperature ($T'_{H}$)
$\Gamma_{B}=\exp\left[\frac{(E-eA_{0})}{T'_{H}}\right]$, the $T'_{H}$ for RBHCS can be calculated as
\begin{eqnarray}
T'_{H}=\frac{m_{\circ}\left[1+\left(\frac{r_\circ}{r_+}\right)^q\right]^{-\frac{p+q}{q}}
\left[1-(p-1)\left(\frac{r_\circ}{r_+}\right)^q\right]}{2\pi\left(r^2_++a^2\right)}\Big[1-\beta\wp\Big].\label{kk}
\end{eqnarray}

The $T'_{H}$ of BH depends upon the mass $m_{\circ}$,
length $r_{\circ}$, BH radius $r_+$, rotation parameter $a$ and  gravity parameter $\beta$.
We obtain the Hawking temperature for RBHCS in Newman-Janis algorithm (Eq. (\ref{22}))
in the absence of quantum gravity parameter (i.e., $\beta=0$).
After substituting the value of $\wp$, we obtain as
\begin{eqnarray}
T'_{H}=\frac{m_{\circ}\left[1+\left(\frac{r_\circ}{r_+}\right)^q\right]^{-\frac{p+q}{q}}
\left[1-(p-1)\left(\frac{r_\circ}{r_+}\right)^q\right]}{2\pi\left(r^2_++a^2\right)}\Big[1-\beta 6\left(m^{2}+\frac{j^{2}}{r^{2}_{+}}\right)\Big]
\end{eqnarray}
where
$m^{2}+\frac{j^{2}}{r^{2}_{+}}$ represents the radiated particles kinetic energy. The kinetic energy is approximated as $\omega$ for residual mass. Quantum corrections accelerate the increase in temperature during the radiation phenomenon.
 These corrections make the radiation cease at some particular Hawking temperature, leaving the remnant mass. When this consideration holds, the temperature stops rising \cite{R23}
\begin{equation}
(M-dM)(1+\beta\wp)\approx M,
\end{equation}
where $\omega=dM,~~\beta_{0}(\frac{1}{M_{Planks}})=\beta $ and $M_{Planks}=\omega$. Here, $\beta_0$ and $M_{Planks}$ are represent the dimensionless parameter and Planck
mass and 
quantum gravity effects for $\beta_0 < 10^{5}$ in \cite{R24, R25}. 

\section{Graphical Stability Analysis}
The current section gives the analysis of $T'_{H}$ of RBHCS in Newman-Janis algorithm. We analyze the physical
significance of  $T'_{H}$ versus horizon $r_{+}$
and study the physical and stable condition of the corresponding RBHCS.

\textbf{Fig. 2}: (i) interpretation of $T'_{H}$ for fixed $m_{\circ}=50$, $r_{\circ}=10, a=0.1, \wp=10$,
$p=3, q=2$ and varying values of correction parameter $\beta$ in the range $0\leq r_{+}\leq15$. Firstly,
the $T'_{H}$ slowly increases and after getting a maximum height the
$T'_{H}$ gradually reductions with the increasing $r_+$ and gets an asymptotically
flat state as $r_{+}\rightarrow\infty$. This behavior comprise the stable state of BH.

(ii) shows the interpretation of $T'_{H}$ via $ r_{+}$ for fixed $m_{\circ}=50$, $r_{\circ}=10, \beta=500, \wp=10$,
$p=3, q=2$ and varying the rotation parameter $a$ values in the range $0\leq r_{+}\leq12$.
The $T'_{H}$ exponentially increases and at a high $T'_{H}$, it attains
a maximum height and then decreases as $T'_{H}\rightarrow0$. We can observe
that as we increase the rotation parameter values then the $T'_{H}$ goes on decreasing.
The decreasing $T'_{H}$ with increasing horizon depicts the physical state of BH.

\begin{center}
\includegraphics[width=7cm]{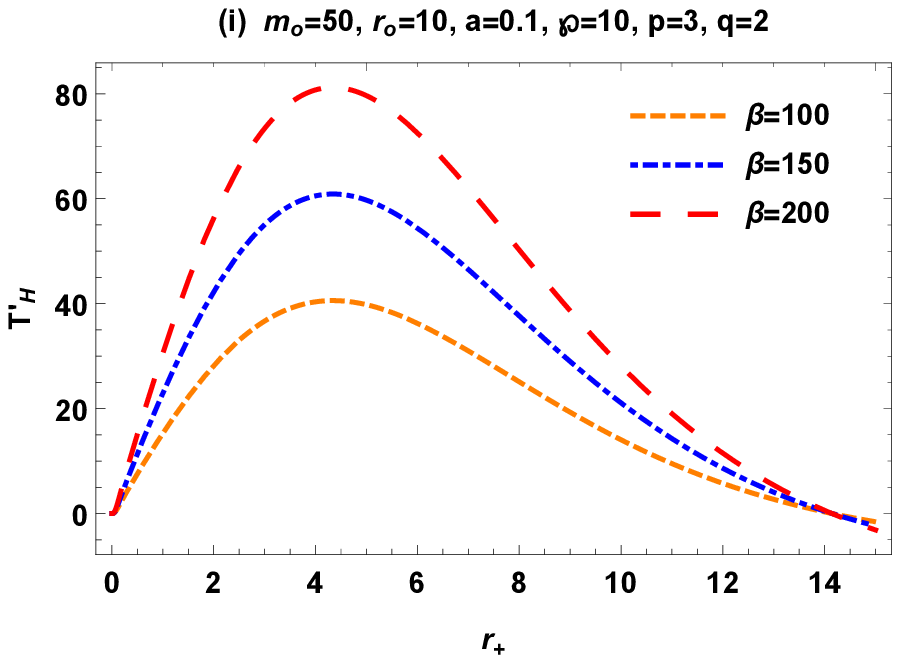}\includegraphics[width=7cm]{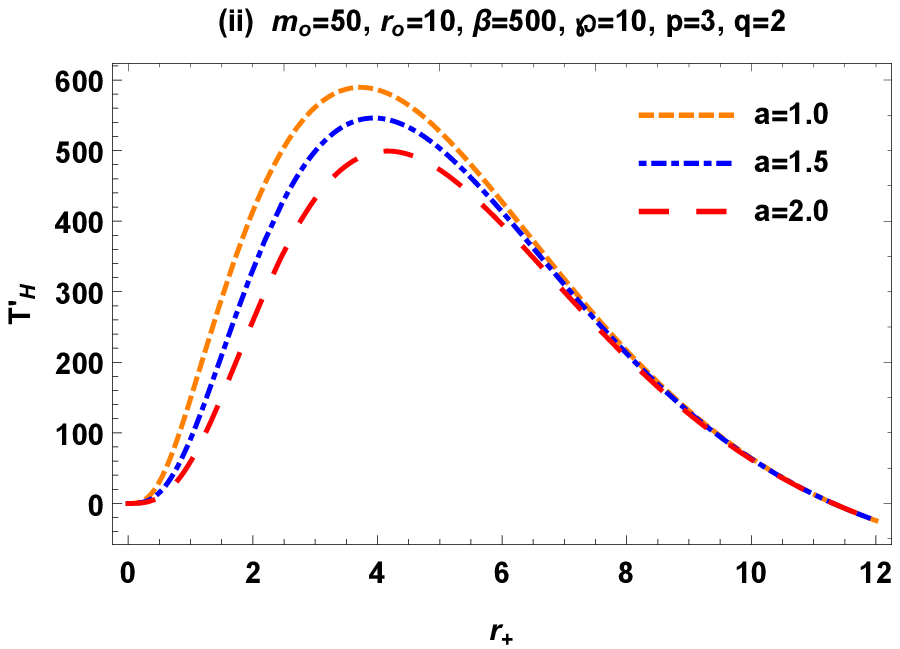}\\
{Figure 2: $T'_{H}$ versus $r_{+}$.}
\end{center}

\section{Corrected Entropy for Regular Black Holes with Cosmic Strings with Rotation Parameter}
This section investigates the entropy corrections for RBHCS with rotation parameter. Banerjee et al, \cite{R3,R4, R5} have computed the corrections of Hawking temperature as well as entropy by considering the back-reaction effects through null
geodesic technique. We derive the entropy corrections for rotating RBHCS by taking into account the Bekenstein-Hawking entropy formula for leading (first) order corrections (Pradhan, 2017).
We investigate the logarithmic entropy corrections by considering the corrected temperature $T'_{H}$ and standard entropy $\mathbb{S}_{o}$ for rotating RBHCS.
The entropy corrections can be derived by the given formula 
\begin{equation}
\mathbb{S}=\mathbb{S}_{o}-\frac{1}{2}\ln\Big|T_H^2, \mathbb{S}_{o}\Big|+...~.\label{vv}
\end{equation}
The standard entropy for rotating RBHCS can be computed by the following formula
\begin{equation}
\mathbb{S}_{o}=\frac{A_+}{4},
\end{equation}
where
\begin{eqnarray}
A_+&=&\int_{0}^{2\pi}\int_{0}^{\pi}\sqrt{g_{\theta\theta}g_{\phi\phi}}d\theta d\phi,\nonumber\\
&=&\frac{2\pi\zeta\Big[\Big(r_{+}^2+a^2\Big)^2-a^2r_{+}M\zeta^2\Big]}{\Big(r_{+}^2+a^2\Big)}.
\end{eqnarray}
So the standard entropy term for rotating RBHCS can be calculated as
\begin{eqnarray}
\mathbb{S}_{o}&=&\frac{\pi\zeta\Big[\Big(r_{+}^2+a^2\Big)^2-a^2r_{+}M\zeta^2\Big]}{2\Big(r_{+}^2+a^2\Big)},\nonumber\\
&=&\frac{\pi\zeta\Big[\Big(r_{+}^2+a^2\Big)^2-a^2r_{+}m_{\circ}\left[1+\left(\frac{r_\circ}{r_+}\right)^q\right]^{-\frac{p}{q}}\zeta^2\Big]}{2\Big(r_{+}^2+a^2\Big)}.\label{v1}
\end{eqnarray}
After inserting the values from Eq. (\ref{kk}) and (\ref{v1}) into Eq. (\ref{vv}),
we get the entropy corrections as follows
\begin{eqnarray}
\mathbb{S}&=&\frac{\pi\zeta\Big[\Big(r_{+}^2+a^2\Big)^2-a^2r_{+}m_{\circ}\left[1+\left(\frac{r_\circ}{r_+}\right)^q\right]^{-\frac{p}{q}}\zeta^2\Big]}{2\Big(r_{+}^2+a^2\Big)}\nonumber\\
&-&\frac{1}{2}\ln\left|\frac{\left[m_{\circ}\left[1+\left(\frac{r_\circ}{r_+}\right)^q\right]^{-\frac{p+q}{q}}
\left(1-(p-1)\left(\frac{r_\circ}{r_+}\right)^q\right)\right]^2\Xi\Big[1-\beta\wp\Big]^2}{8\pi\left(r^2_++a^2\right)^3}\right|+...,\label{b2}
\end{eqnarray}
where
\begin{equation}
\Xi=\zeta\Big[\Big(r_{+}^2+a^2\Big)^2-a^2r_{+}m_{\circ}\left[1+\left(\frac{r_\circ}{r_+}\right)^q\right]^{-\frac{p}{q}}\zeta^2\Big].\nonumber
\end{equation}
The Eq. (\ref{b2}) gives the entropy corrections for rotating RBHCS.

\section{Entropy Analysis}
This section gives the analysis of corrected entropy $S$ versus horizon $r_+$ for RBHCS in Newman-Janis algorithm. We discuss the effects of spin parameter $a$  and correction parameter $\beta$ on corrected entropy $S$ with the help of graphical interpretation for fixed values of parameters $p=3$ and $q=2$.

\textbf{Fig. 3}: (i) depicts the graphical interpretation of $S$ for fixed $m_0=10$, $r_0=0.1$, $a=5$, $\wp= 0.1=\zeta$ and varying values of correction parameter $\beta$ in the range $0\leq r_{+}\leq 5$. Firstly,
the $S$ slowly decreases and after getting a lowest value the
$S$ gradually increases for increasing values of $r_+$. This behavior comprise the stable state of corrected entropy.

(ii) shows the behavior of $S$ via $ r_{+}$ for fixed $m_0=5$, $r_0=0.1$, $\beta=0.5$, $\wp= 0.1=\zeta$ and varying values of the rotation parameter $a$ in the range $0\leq r_{+}\leq 10$.
Initially, the $S$ exponentially decreases and after a minimum height
it goes on increasing and shows convergence. We can observe
from plot that with the increasing values of $a$ entropy also increases.\\
From both plots, we can observe that the entropy shows physical form for greater/smaller values of correction parameter. So, we conclude that the BH entropy shows stable form under quantum gravity effects.
\begin{center}
\includegraphics[width=7cm]{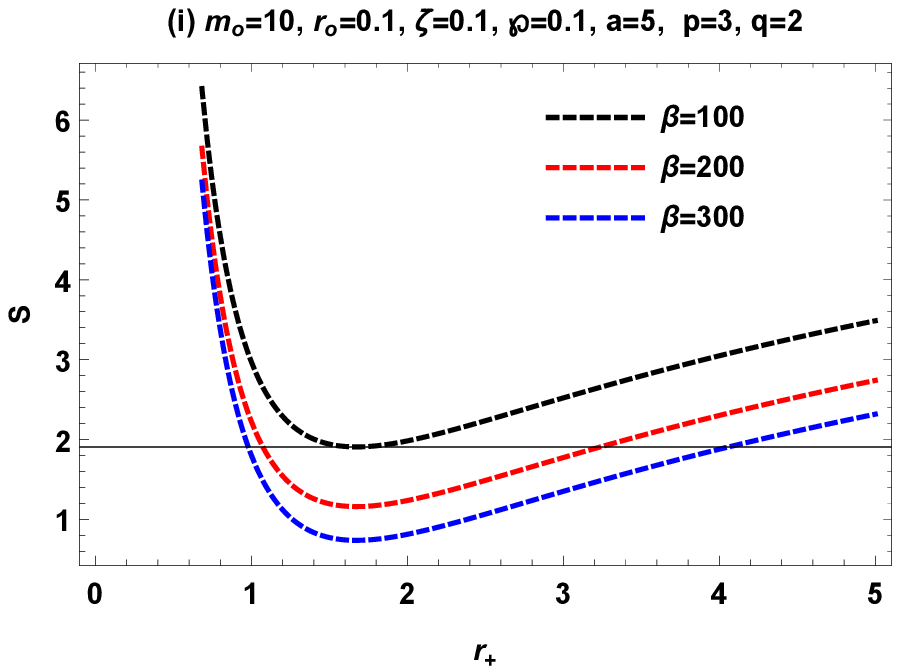}\includegraphics[width=7cm]{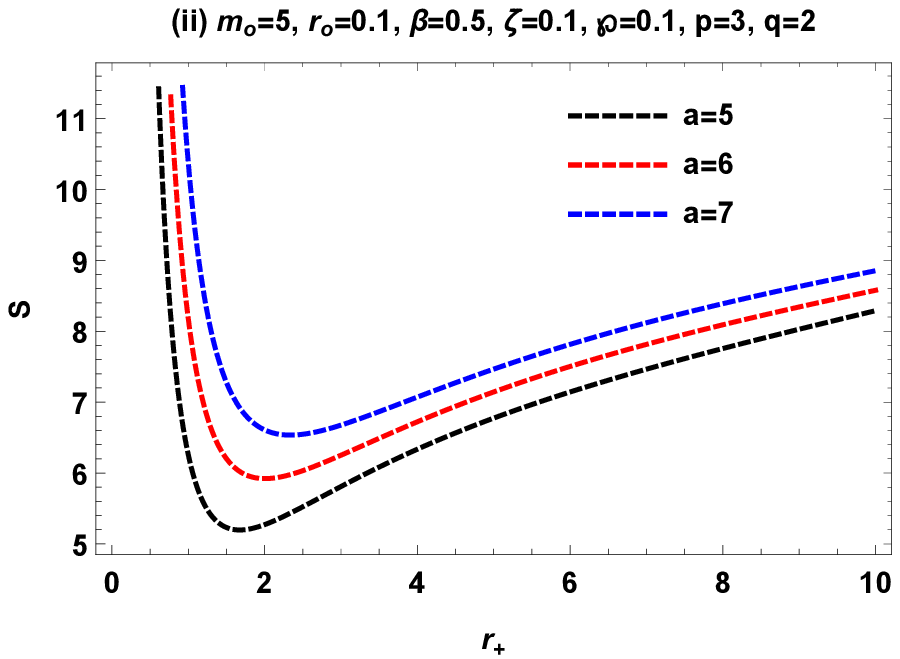}\\
{Figure 3: (i) $S$ via $r_{+}$.}
\end{center}
\textbf{Fig. 4}: gives the behavior of entropy via horizon for $m_0=1$, $a=1.5$, $\beta=50$, $\wp= 0.1=\zeta$ and for different values of $r_0$ in the range $0\leq r_{+}\leq 8$. From our plot, we can observe entropy is unstable for the values of $r_0\geqslant1$. So, Fig. \textbf{4} shows the unstable form for BH entropy under the influence of quantum gravity effects.

\begin{center}
\includegraphics[width=7cm]{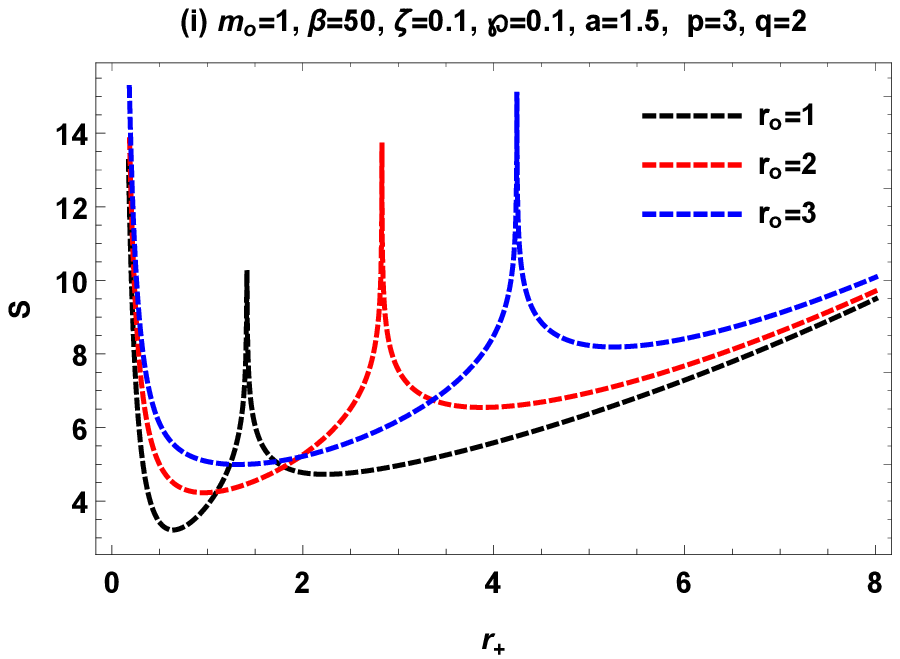}\\
{Figure 4: $S$ via $r_{+}$.} 
\end{center}

\section{Summary and Discussion}
In this article, we have studied the RBHCS solution in the background of rotation parameter by considering the Newman-Janis method. By considering the spin coupling
$(a\rightarrow 0)$ in the Eq. (\ref{01}), we get the RBHCS solution without rotation parameter.
The RBHCS solution in Newman-Janis algorithm is quite different from the BH solution in GR.
Moreover, we have investigated the thermodynamical property (i.e., Hawking temperature)
for RBHCS in the presence of rotation parameter. This temperature depends upon the mass $m_{\circ}$,
length $r_{\circ}$, BH radius $r_+$ and rotation parameter $a$. We get the Hawking temperature
for RBHCS in the absence of spin parameter (i.e., $a=0$).
It has worth to mention here that, by considering $m_\circ=m, r_{\circ}=0=a$ at
$r_+\thickapprox2m$, we recover the Schwarzschild Hawking temperature $T_{Sch}=1/8\pi m$.
The RBHCS in the context of spin parameter are colder than the Schwarzschild BH.
We have investigated this phenomenon graphically. We have concluded that the temperature
in plot (ii) is higher than in plot (i), when we consider $m_\circ=m, r_{\circ}=0=a$ in
Eq. (\ref{22}). We have studied the physical significance of temperature $T_{H}$ via horizon
$r_{+}$ in the presence/absence of spin parameter and analyzed the physical and stable condition
of the corresponding RBHCS. Based on the Hawking phenomenon, as the temperature rises and more radiation is
released, the BH radius decreases. The stability of BH is defined by this physical phenomenon.
Furthermore, we have investigated the quantum corrected temperature for RBHCS in Newman-Janis algorithm by incorporating GUP effects.
We observed that in the absence of quantum gravity parameter (i.e., $\beta=0$), we recover
the Hawking temperature for RBHCS in Newman-Janis algorithm (Eq. (\ref{22})). We have also analyzed
the graphical interpretation of corrected temperature $T'_{H}$ versus $r_{+}$ and studied the
stable condition of RBHCS in Newman-Janis algorithm under the quantum gravity influence. 

Finally, the corrected entropy for RBHCS with rotation parameter is investigated. When the gravity parameter influences are ignored and we have obtained the absolute entropy of RBHCS. In future, we have computed the corrected temperature associated
with the emission rate of spin–1 bosons from a black rings metric and wormholes metric.

\end{document}